\documentclass[twocolumn,showpacs,amsmath,amssymb,superscriptaddress]{revtex4}

\usepackage{epsfig,psfrag}
\usepackage{dcolumn}% Align table columns on decimal point
\usepackage{bm}% bold math
\usepackage{graphicx}
\usepackage{color}
\newcommand{\beq}{\begin{equation}}
\newcommand{\eeq}{\end{equation}}
\newcommand{\beqr}{\begin{eqnarray}}
\newcommand{\eeqr}{\end{eqnarray}}
\newcommand{\e}{{\epsilon}}

\def\tb{{\tilde{b}}}
\def\tA{{\tilde{A}}}

\def\bA{{\mathbf A}}

\def\bsig{{\mathbf \sigma}}

\def\bn{{\mathbf n}}

\def\bq{{\mathbf q}}
\def\hbq{{\hat{\mathbf q}}}

\def\bb{{\mathbf b}}

\def\bJ{{\mathbf J}}

\def\ba{{\mathbf a}}

\def\half{{1\over2}}

\def\eqa{\begin{eqnarray}}
\def\eea{\end{eqnarray}}
\def\cA{{{\cal A}}}

\def\cL{{{\cal L}}}

\def\cS{{{\cal S}}}

\def\ssc{Sol. State. Commun.}
\def\njp{New Jour. Phys.}

\def\epl{Europhys. Lett.}

\begin{document}

\title{Quantum Hall to Insulator Transition in the Bilayer Quantum Hall Ferromagnet}
\author{Ganpathy Murthy} 
\affiliation{Department of Physics and Astronomy,
University of Kentucky, Lexington KY 40506-0055} 
\author{Subir Sachdev}
\affiliation{Department of Physics, Harvard University, Cambridge MA 02138}
\date{\today}
\begin{abstract}
We describe a new phase transition of the bilayer quantum Hall
ferromagnet at filling fraction $\nu = 1$.  In the presence of static
disorder (modeled by a periodic potential), bosonic $S=1/2$ spinons
can undergo a superfluid-insulator transition while preserving the
ferromagnetic order. The Mott insulating phase has an emergent U(1)
photon, and the transition is between Higgs and Coulomb phases of this
photon.  Physical consequences for charge and counterflow
conductivity, and for interlayer tunneling conductance in the presence
of quenched disorder are discussed.
\end{abstract}
\vskip 1cm 
\maketitle
The quantum Hall effects embody new states of matter, in which
two-dimensional electron systems in a perpendicular magnetic field $B$
are incompressible, and exhibit excitations with fractional charge
and statistics \cite{QHE}.

Quantum Hall systems with an internal degree of freedom, such as spin,
layer index, or valley index, are richer still
\cite{multicomp-review}. Because electron or hole excitations are
prohibitive in energy, the low energy states can be characterized by
orientation of the vector $\bn$, denoting the spin (or pseudospin in
the layer index). Exchange interactions lead to ferromagnetism, hence
such systems are quantum Hall ferromagnets (QHFM's). Of central
importance is the spin-charge relation \cite{skyrmion,moon} in the
lowest Landau level (LLL), which expresses the Coulomb charge and
current $(J^{0},\bJ)=J^{\mu}$ due to a varying configuration of $\bn$
\beq
J^{\mu}_{S}={e\over8\pi}\epsilon^{\mu\nu\lambda} \bn\cdot(\partial_{\nu}\bn\times\partial_{\lambda}\bn)
\eeq
The charge carriers in QHFM's are {\it spin-textures},
characterized by a topological number. In single-layer
$\nu=1$ systems with spin, they are fermionic
skyrmions/antiskyrmions\cite{skyrmion} with charge $\pm e$, while in
bilayer systems they are quartonic merons/antimerons\cite{moon} with
charge $\pm{e/2}$.

While our general framework applies to all quantum Hall ferromagnets,
in this paper we focus on the bilayer quantum Hall system
\cite{multicomp-review,eisenstein} at total filling factor $\nu_T=1$,
with an extremely small, but nonzero, tunneling amplitude $h$ between
the two layers. We will assume that real spin is frozen (which may not
entirely be valid \cite{spin-in-bilayer}), and spin/peudospin for us
will be synonymous with the layer index. Each layer is at
half-filling. When the separation between the layers $d$ is of the
order of a magnetic length $l_0=\sqrt{{hc/ eB}}$, the system is an
incompressible quantum Hall state \cite{murphy}. At large enough
$d/l_0$ the system splits up into two weakly interacting,
compressible, presumably $\nu=\half$ Fermi-liquid-like systems
\cite{murphy}.

Despite almost two decades of theory
\cite{fertig89,wen,moon,balents,kimetal,wang} and experiments
\cite{spielman1,spielman2}, important aspects of the
small $d/l_0$ phase are not understood. Theoretically, the small
$d/l_0$ QHFM breaks symmetry at $T=0$ even in the absence of an
interlayer tunneling\cite{wen,moon} (spontaneous interlayer coherence)
leading to a Goldstone mode\cite{fertig89}. This state can also be
regarded as an excitonic superfluid, and should exhibit a
Josephson-like effect, with a finite interlayer current flowing at
strictly zero interlayer bias voltage\cite{wen,moon}. While there is a peak in
the interlayer tunneling conductance $G$ at zero
bias \cite{spielman1}, the peak has finite width, implying some
intrinsic dissipation. Theoretically, interlayer tunneling should take
place only within a Josephson length of the contacts
\cite{moon,rossi-macd}. Thus $G$ should be proportional to the length
of the contacts, while experimentally it is seen to be proportional to
the area \cite{area-vs-perimeter}. Theoretically, there should be a
$T>0$ Kosterlitz-Thouless transition at which the superfluid stiffness
has a universal jump. Experimentally\cite{KT-transition}, the
zero-bias value of $G$ (the closest analog to the superfluid
stiffness) vanishes roughly as $(T_c-T)^3$ at the transition. It is
believed\cite{balents,herb-deconfine,sheng,huse} that quenched
disorder is ultimately responsible for these discrepancies, though a
detailed understanding is lacking.

Together with H. A. Fertig, one of us has proposed a
model\cite{coh-network} where a ``coherence network'' forms due to the
nonperturbative effects of disorder\cite{efros}. This model is
consistent with several aspects of the experiments, notably the
tunneling conductance going as the area\cite{area-vs-perimeter} rather
than the length of the sample, but it is classical.

In this paper we argue for a quantum phase transition to an insulating
zero-temperature phase of QHFM, which could possibly impact on the
experimental issues discussed above. We construct a quantum low energy
model at $T=0$ (neglecting electron and hole excitations and keeping
only smooth configurations of $\bn$), and mimic the nonperturbative
effects of disorder\cite{efros} by putting the system on a square
lattice. We start with the imaginary time Lagrangian for a
two-component quantum Hall system as described by Lee and
Kane(LK)\cite{lee-kane}
\beqr
{\cal{L}}_{LK}=&{i\over4\pi}\e_{\mu\nu\lambda}a_{\mu}\partial_{\nu}a_{\lambda}+\bar{\Psi}_s\big(\partial_0-i(a_0-eA_0)\big)\Psi_s\nonumber\\
&+{1\over2m}|\big(\nabla-i(\ba-e\bA)\big)\Psi_s|^2+{u\over2}\big(\bar{\Psi}_s\Psi_s-\rho_0\big)^2
\eeqr
Here the $\Psi_{s}$ are the composite boson (CB) fields\cite{ZHK},
obtained by attaching one unit of statistical flux to the
electron. The gauge field $a_{\mu}$ with its Chern-Simons term
implements this transformation, while the physical (Coulomb) charge
$-e$ of the CB leads to minimal coupling to the external
electromagnetic field $eA_{\mu}$. When the filling is exactly $\nu=1$
($\rho_0=1/2\pi l_0^2$), the statistical and external fields cancel
each other on average, and the CB's can Bose condense, which leads to
the QH phase with a quantized Hall conductance\cite{ZHK}. One now
decomposes\cite{lee-kane} the CB field into a single-component charged
Higgs field $\Phi$, and a neutral two-component unit length spinor
$z_s$: $\Psi_{s}=\Phi z_{s}$ ($\bar{z}_sz_s=1$). The spin vector
emerges as $\bn=\bar{z}\bsig z$. We parameterize the spin sector by
the variables $\zeta=n_z$ and the angle $\theta$ of the $xy$ component
of spin, and also add the planar anisotropy energy $\Gamma$. After
some manipulations, one obtains\cite{lee-kane}
\beqr
&{i\over4\pi}\e_{\mu\nu\lambda}a_{\mu}\partial_{\nu}a_{\lambda}+\bar{\Phi}\big(\partial_0-i(a_0+b_0-eA_0)\big)\Phi\nonumber\\
&+{1\over2m}|\big(\nabla-i(\ba+\bb-e\bA)\big)\Phi|^2
+{\rho_0\over2m}\big(\nabla\bn\big)^2+{\Gamma\over2}n_z^2
\label{LK}\eeqr
Note that the coupling between the Higgs and spin sectors is only via the gauge-like field
\beq
b_{\mu}=i\bar{z}\partial_{\mu}z=\half\zeta\partial_{\mu}\theta\ \ \ \ \ \ \ \ J^{\mu}_{S}={e\over4\pi}\epsilon^{\mu\nu\lambda}\partial_{\nu}b_{\lambda}
\eeq

The idea is to allow the periodic potential to drive the Higgs
field through a Higgs-Mott Insulator transition, while $\bn$
remains ordered. Such a transition without a gauge field
is known to exist for two-component bosons on
lattices\cite{BH-MFT}. The Higgs transition in
single-component quantum Hall systems has been studied previously in
the large-$N$
approximation\cite{Higgs-CS-large-N},
and yields a second-order quantum phase transition at which the Hall
conductance changes discontinuously. We will use the same approach,
and assume that our transition is second-order, though the question of
the order remains open for the physically relevant case  $N=1$.

Since the transition is at fixed Higgs density, it is described by a
relativistic effective theory\cite{bose-hubbard}. 
\beqr
&\cL=|\big(\partial_{\mu}-i\cA_{\mu}\big)\Phi|^2+M^2|\Phi|^2+\lambda|\Phi|^4+{i\over4\pi}\e_{\mu\nu\lambda}a_{\mu}\partial_{\nu}a_{\lambda}+\nonumber\\
&{i\rho_0\over2}\zeta\dot{\theta}+{K\over2}\big((1-\zeta^2)(\nabla\theta)^2+{(\nabla\zeta)^2\over(1-\zeta^2)}\big)+{\Gamma\over2}\zeta^2-h\cos{\theta}
\eeqr
Here we have defined $\cA_{\mu}=a_\mu+b_\mu-eA_\mu$, and allowed the
spin stiffness $K$ to be renormalized down to its lowest Landau level
(LLL) value (as obtained by Moon et al\cite{moon}). We have also
introduced the interlayer tunneling $h$, which we will take to be much
smaller than any other energy scale, and will therefore neglect unless
it is essential for some physical quantity. Note that we have used the
$\bar{\Phi}\Phi\bar{z}\partial_0z$ term of the LK action
(Eq. (\ref{LK})) with $\langle\bar{\Phi}\Phi\rangle=\rho_0$ to obtain
the $\zeta\dot{\theta}$ term in the lagrangian. This is not only the
most relevant time-derivative term, but has the correct dynamics for
the density-density correlations due to spin-textures to be of order
$q^4$ at small $q$, a requirement of being in the LLL. The higher time
derivative terms coming from integrating out high-energy modes will
appear in such a combination as to not violate the LLL property. In
this LLL QHFM, the $2\times2$ correlator of $\langle
b_{\mu}b_{\nu}\rangle$ (with $b_0(k)$ really standing for
$\tb_0(k)=b_0(k)-{\omega\over q^2}\bq\cdot\bb(k)$, see below), for small $q,\omega$ is
\beq
G_b^{(0)}(k)=\left[ \begin{array}{cc} D & 0 \\ 0 & Fq^2 \ \end{array}\right]
\label{bb-bare-corr}\eeq
where $D,\ F$ are constants.

Now we integrate out $\Phi$ to obtain 
\beq
\cS_{eff}={\cS}_{CS}+\int dk \big(\cA_{\mu}(k)\Pi^{\mu\nu}_{\Phi}(k)\cA_{\nu}(-k)\big)
\eeq
Here $k=(\omega,\bq)$ and the polarization tensor $\Pi_{\Phi}$ has the gauge-invariant form
\beq
\Pi^{\mu\nu}_\Phi(k)={f(k^2)\over 4\pi k^2}(k^2\delta^{\mu\nu}-k^{\mu}k^{\nu})
\eeq
In the Higgs phase $f$ is a constant, while in the Mott phase $f\simeq
k^2$ for $k\ll M$.  Going to Coulomb gauge for $a_\mu$
($\nabla\cdot\ba=0$), one can rewrite the gauge action in a $2\times2$
matrix form acting on $a_0(k)$ and $a_T(k)=i{\hbq}\times\ba(k)$, where
$\hbq=\bq/q$. We will make a similar decomposition for $b_\mu$ and
$A_{\mu}$. Define $\tb_0(k)=b_0(k)-{i\omega\over q}b_L(k)$, with
$b_T(k)=i{\hbq}\times\bb(k)$ and $b_L(k)=-i\hbq\cdot\bb(k)$. Then the
entries corresponding to $a_0$ and $a_T$ are $\tb_0$ and $b_T$. In
this $2\times2$ language we have
\beqr
\Pi_{\Phi}=& {1\over4\pi}\left[ \begin{array}{cc} f{q^2\over k^2} & 0 \\ 0 & f \ \end{array}\right]\ \ \ \ 
\Pi_{CS}=&{1\over4\pi} \left[ \begin{array}{cc} 0 & {iq} \\ {iq} & 0 \ \end{array}\right]
\eeqr

To proceed, we define  $\Pi=\Pi_{\Phi}+\Pi_{CS}$ and integrate out the gauge
fields to  obtain an effective action for ${\cal{B}}_0=\tb_0-e\tA_0$ and ${\cal{B}}_T=b_T-eA_T$: 
\beq
\int dk \left(\begin{array}{cc} {\cal{B}}_0(k) & {\cal{B}}_T(k) \end{array}\right)[\Pi_{\Phi}-\Pi_{\Phi}\Pi^{-1}\Pi_{\Phi}]\left(\begin{array}{c} {\cal{B}}_0(-k) \\ {\cal{B}}_T(-k) \end{array} \right)
\eeq
where the matrix $Q=\Pi_{\Phi}-\Pi_{\Phi}\Pi^{-1}\Pi_{\Phi}$ has the form
\beq
Q={1\over4\pi}\left[\begin{array}{cc} {fq^2\over f^2+k^2} & {iqf^2\over f^2+k^2}\\  {iqf^2\over f^2+k^2} & {f k^2\over f^2+k^2} \end{array}\right]
\eeq
In order to obtain the electromagnetic response, one should now
integrate out $b_{\mu}$, that is, $\zeta,\ \theta$.  We take account
of the $b_\mu Q_{\mu\nu} b_{\nu}$ term in the random phase
approximation (RPA) to obtain the ``full'' correlator (as opposed to
the ``bare'' correlator of Eq. (\ref{bb-bare-corr})).
\beq
\langle b_{\mu}(k)b_{\nu}(-k)\rangle = G^{RPA}_{b,\mu\nu}=\big((G_b^{(0)})^{-1}+Q\big)^{-1}_{\mu\nu}
\label{bb-full-corr}\eeq
We now integrate out $b$ to obtain the final effective action for the
electromagnetic potential $A_{\mu}$, which takes the form 
\beq
S_{eff}[A_{\mu}]=\int dk {e^2\over4\pi} A_{\mu}(k)\big(Q-QG_b^{RPA}Q)_{\mu\nu} A_{\nu}(-k)
\eeq
Now we can read off the conductivity matrix from 
\beq
P_{\mu\nu}=4\pi\big(Q-QG_b^{RPA}Q\big)=4\pi\big(G_b^{(0)}+Q^{-1}\big)^{-1}_{\mu\nu}
\eeq
Going to the real frequency domain, we obtain
\beqr
Re(\sigma_{xx}(q,\omega))=&{e^2\over2\pi} {\omega\over q^2} Im(P_{00})\\
\sigma_{xy}(q,\omega)=&{e^2\over2\pi} {1\over q} P_{01}
\eeqr
In the Higgs phase, $Re(\sigma_{xx})$ vanishes below the gap $f$ and
$\sigma_{xy}$ approaches the quantized value of $e^2/2\pi$ for
$\omega,\ q\ll f$ denoting a quantized Hall
phase. In the Mott phase the entire conductivity matrix vanishes for
$\omega,\ q\ll M$, denoting an insulating phase. At the critical point
$f={\pi\over8}\sqrt{q^2-\omega^2}$, and the system has  the critical
conductivities
\beq
\sigma^*_{xx}(0,\omega)={e^2\over2\pi} {\pi/8\over1+(\pi/8)^2}\ \ \ \ \ \ \sigma^*_{xy}={\pi\over8}\sigma^*_{xx}
\eeq

Going back to Eq. (\ref{bb-full-corr}), one can find propagating
charge modes as poles of $G_b^{RPA}$. In the Higgs phase, one finds a
propagating mode with a dispersion
\beq
\omega=\sqrt{q^2(1+Df)+f^2/(1+Ffq^2)}
\eeq
In the Mott phase, no propagating charge modes exist (there are only
branch cuts in the charge correlator).

Let us now turn to the spin sector.  The most interesting quantity is
the self-energy matrix (in $\zeta,\theta$ space) $\Sigma_{ab}$ near
the critical point, which can be related to measurable quantities.  For example
$Im(\Sigma_{\theta\theta})$ is related to the dissipative counterflow
conductivity via
\beq
Re(\sigma_{CF}(\bq,\omega))\simeq K^2q^2 {Im(\Sigma_{\theta\theta}(\bq,\omega))\over\omega[(\omega-c_{\theta}q)^2+(Im(\Sigma_{\theta\theta}))^2]}
\eeq
and to the interlayer conductance via
\beq
G_{interlayer}=2\rho_0\omega{Im(\Sigma_{\theta\theta}(\bq,\omega))\over[(\omega-c_{\theta}q)^2+
(Im(\Sigma_{\theta\theta}))^2]}
\eeq
where we have absorbed the real part of $\Sigma_{ab}$ into a
renormalization of the spin-wave velocity.

In the Higgs phase we find for small $q$ close to threshold $ \omega-f\ll f$
\beqr
Im\big(\Sigma_{\theta\theta}\big)(\omega,q)&\simeq q^2f^2 (K(\omega -f)^2+h) \Theta(\omega-f)\nonumber\\
Im\big((\Sigma_{\zeta\zeta}\big)(\omega,q)&\simeq q^2f^2 (K(\omega -f)^2+\Gamma) \Theta(\omega-f)\nonumber\\
Im\big(\Sigma_{\zeta\theta}\big)(\omega,q)&\simeq q^2f^2\Theta(\omega-f)
\eeqr
where we define the off-diagonal term in the quadratic term of the
effective action of $\zeta,\theta$ in real frequency as
$(1+\Sigma_{\zeta\theta})i\omega\rho_0/2$.  The main features are the
presence of a threshold frequency (the gap to charge excitations)
which vanishes linearly as the Higgs field approaches criticality
$f\to0$, as well as a coupling which also vanishes as a power of
$f$. Note the qualitative difference made in
$Im\big(\Sigma_{\theta\theta}\big)$ by the presence of $h$.

At the large-$N$ critical point we find
\beqr
Im{\Sigma_{\theta\theta}}&\simeq (\omega^7,q^7):\ \ \ \ h=0\\
&\simeq (\omega^5,q^5):\ \ \ \ h\ne0
\eeqr
Here we see that powers of $f$ have been replaced by powers of
$\omega,q$, with one extra power appearing due to the branch cut.

Let us discuss finite-$T$ properties briefly.  Starting in the Higgs
phase, if one goes to nozero $T\gg f$ the system is quantum
critical\cite{subir-book} and we can expect to see the above behavior
with $\omega\to T$. Also, in the presence of quenched disorder, we can
expect (in addition to the activated contribution due to the tunneling
of merons\cite{coh-network}) a power-law contribution in $T$ to all
physical quantities in the quantum critical regime.

Consider now the qualitative behavior of the interlayer
conductance $G$ for
$T\gg h$, believed to be true of experimental samples at
millikelvin temperatures. For $T\gg f$ we expect the interlayer
tunneling to be incoherent, which is consistent with the finite width
of the zero-bias peak seen in experiments. With quenched disorder
there may be a Higgs glass phase (see below) with gapless charge
fluctuations, which would imply incoherent interlayer tunneling for
any nonzero $T$.

Finally, deep in the Mott phase, there is again a threshold frequency
$\omega_{th}(\bq)=\sqrt{4M^2+q^2}$.
\beq
Im\Sigma_{\theta\theta}(\omega,q)\approx (\omega-\omega_{th})^4 q^2 \Theta(\omega-\omega_{th})
\eeq
The addition of the long-range Coulomb interaction to the action makes
no qualitative difference to the above.

Let us now comment on previous related work. Some authors have argued
for an $XY$ spin-glass phase at sufficient
disorder\cite{spin-glass-QHFM,coh-network} but these arguments are for
the classical model. There are also proposals that the ground state at
$T=0$ is a gauge-glass\cite{balents,sheng} which is still a quantum
Hall phase, but with power-law $XY$ order. Other authors have argued
for a {\it spontaneous} breaking of translation
invariance\cite{yang-WC}: Like the quantum Hall-Wigner
crystal transition this likely occurs at large
imbalance\cite{yang-WC}. Finally, some authors have argued for a
translation-invariant QH phase with no long-range order in
$\bn$\cite{caldeira}.  In our model, we assume that the
$XY$ ferromagnetism is robust across the Higgs transition which is
driven by a periodic potential (a proxy for disorder), which differs
from all the above proposals.

In the presence of static disorder, a key feature of our model is that
the dissipation arises not directly from disorder coupled to the $XY$
order parameter\cite{localization-in-QHFM}, but rather from the
disorder inducing a phase transition which creates a phase with {\it
dynamical} low-energy spinon and charge excitations coupled to the
$XY$ modes. We call this phase a Higgs glass, in analogy to the Bose
glass\cite{bose-hubbard} and gauge glass\cite{balents,sheng}
phases. The Higgs glass differs from the gauge glass in having {\it
dynamical, gapless gauge fluctuations}.  We expect that there will be
an imperfect charge-flux relation in the Higgs glass phase, which will
allow us to infer gapless charge fluctuations with perhaps a vanishing
density of states at vanishing energy.

In conclusion, we have studied a lowest Landau level model for the
$\nu=1$ bilayer quantum Hall ferromagnet which displays a
Higgs$\to$Mott transtion of spinons in the presence of a periodic
potential.  Our understanding of the Mott phase of the spinons, which
is also a ferromagnetic insulator, remains incomplete. The standard
expectation is that the large $d/l_0$ bilayer system is best described
in terms of two species of weakly interacting Composite Fermions
(CF's)
\cite{kimetal}. It is possible that the transition we have
described on the lattice is preempted by a first-order transition
(reverting to second-order with quenched disorder) to a phase
adiabatically connected to two decoupled species of CF's. 

The framework we have sketched should apply to the single layer QHFM
with spin, where the existence of charge fluctuations below the Zeeman
energy is not understood\cite{disorder-in-QHFM1}. 

It is a pleasure for GM to thank Herb Fertig, Ziqiang Wang, Steve
Girvin, and T. Senthil for illuminating discussions, and the Aspen
Center for Physics where some of this work was conceived. GM also
deeply appreciates the hospitality of the Physics Department at
Harvard University, where this work was carried out. We would like to
acknowledge partial support from the NSF under DMR-0703992 (GM) and
DMR-0757145 (SS).


\begin{thebibliography}{99}
\bibitem{QHE} K. von Klitzing, G. Dorda, and 
M. Pepper, \prl {\bf 45}, 494 (1980); D. C. Tsui, H. L. Stormer, and
A. C. Gossard, \prl {\bf 48}, 1559 (1982); R.B.Laughlin, {\it Phys.
Rev. Lett.} {\bf 50}, 1395, (1983); J. K. Jain, {\it Phys.  Rev.
Lett. } {\bf 63}, 199, (1989).
%
\bibitem{multicomp-review} See S. M. Girvin and 
A. H. MacDonald in {\it Perspectives on Quantum Hall Effects}, S. Das
Sarma and A. Pinczuk, Editors (Wiley Interscience, New York, 1997). 
%
%\bibitem{bilayers} L.V. Butov et al., Nature {\bf 417}, 47 (2002); D. Snoke, Science {\bf 298}, 1368 (2002).
%
\bibitem{skyrmion} S.L. Sondhi, A. Karlhede, S.A. Kivelson, and E.H. Rezayi,
Phys. Rev. B {\bf 47}, 16419 (1993).
%
\bibitem{moon} K. Yang, K. Moon, L. Zheng, A. H. MacDonald, 
S. M. Girvin, D. Yoshioka, and S.-C. Zhang, \prl {\bf 72}, 732
(1994); K. Moon, H. Mori, K. Yang, S. M. Girvin, A. H. MacDonald,
L. Zheng, D. Yoshioka, and S.-C. Zhang, Phys. Rev. B {\bf 51}, 5138
(1995).
% 
\bibitem{eisenstein} See J.P. Eisenstein and A.H. MacDonald, Nature {\bf 432}, 691 (2004)
and references therein.
%
\bibitem{spin-in-bilayer} N. Kumada, K. Muraki, K. Hashimoto, 
and Y. Hirayama, \prl {\bf 94}, 096802 (2005); I. B. Spielman,
L. A. Tracy, J. P. Eisenstein, L. N. Pfeiffer, and K. W. West, \prl
{\bf 94}, 076803 (2005); S. Luin, V. Pellegrini, A. Pinczuk, 
B. S. Dennis, L. N. Pfeiffer, anf K. W. West, \prl {\bf 94}, 146804 (2005). 
%
\bibitem{murphy} S. Q. Murphy, J. P. Eisenstein, 
G. S. Boebinger, L. N. Pfeiffer, and K. W. West, \prl {\bf 72}, 728
(1994).
%
\bibitem{fertig89} H.A. Fertig, Phys. Rev. B {\bf 40}, 1087 (1989).
%
\bibitem{wen}X.G. Wen and A. Zee, Phys. Rev. Lett. {\bf 69}, 1811 (1992);
Phys. Rev. B. {\bf 47}, 2265 (1993); Z.F. Ezawa and A. Iwazaki, Phys. Rev. B
{\bf 47}, 7295 (1993).
%
\bibitem{balents} L. Balents and L. Radzihovsky, 
Phys. Rev. Lett. {\bf 86}, 1825 (2001); 
A. Stern, S.M. Girvin, A.H. MacDonald, and Ning Ma,
Phys. Rev. Lett. {\bf 86}, 1829 (2001).
%
\bibitem{kimetal} Y. B. Kim, C. Nayak, E. Demler, N. Read, and 
S. Das Sarma, \prb {\bf 63}, 205315 (2001).
%
\bibitem{wang} Z. Wang, Phys.  Rev. Lett. {\bf  92}, 136803 (2004).
%
%\bibitem{boebinger} G. Boebinger, H.W. Jiang, L.N. Pfeiffer, and K.W. West,
%Phys. Rev. Lett. {\bf 64}, 235 (1990).
%
\bibitem{spielman1} I. B. Spielman, J. P. Eisenstein, L. N. Pfeiffer, and K. W. West,
Phys. Rev. Lett. {\bf 84}, 5808 (2000).
%
\bibitem{spielman2} I. B. Spielman, J. P. Eisenstein, L. N. Pfeiffer, and
%K. W. West, Phys. Rev. Lett. {\bf 87}, 036803 (2001); M. Kellogg, J.P. Eisenstein, L.N. Pfeiffer, and K.W. West,
Phys. Rev. Lett. {\bf 93}, 036801 (2004); E. Tutuc, M. Shayegan, and D. Huse,  
Phys. Rev. Lett. {\bf 93}, 036802 (2004).
%
\bibitem{rossi-macd} E. Rossi, A. S. N\'u\~nez, and 
A. H. MacDonald, \prl {\bf 95}, 266804 (2005). 
%
\bibitem{area-vs-perimeter} A. D. K. Finck, A. R. Champagne, 
J. P. Eisenstein, L. N. Pfeiffer, and K. W. West, {\it
Bull. Am. Phys. Soc.}, {\bf J37.4} (2008).
%
\bibitem{KT-transition} A. R. Champagne, J. P. Eisenstein, 
L. N. Pfeiffer, and K. W. West, \prl {\bf 100}, 096801 (2008).
%
\bibitem{herb-deconfine} H. A. Fertig, Phys. Rev. Lett. 89, 035703 (2002); 
H.~A. Fertig and Joseph P. Straley, Phys.\ Rev.\ B. {\bf 66},
201402(R) (2002); \prl {\bf 91}, 046806 (2003).
%
\bibitem{sheng} D. Sheng, L. Balents, and Z. Wang, Phys.  Rev. Lett. {\bf  91},
116802 (2003).
%
\bibitem{huse} D.A. Huse, cond-mat/0407452 (2004).
%
\bibitem{coh-network} H. A. Fertig and G. Murthy, \prl {\bf 95}, 
156802 (2005); \ssc {\bf 140}, 83 (206).
%
\bibitem{efros} A. L. Efros, Sol. St. Commun. {\bf 65}, 1281 (1988);
A. L. Efros, F. G. Pikus, and V. G. Burnett, \prb {\bf 47}, 2233
(1993).
%
\bibitem{ZHK} S.-C. Zhang, H. Hansson, and S. A. Kivelson, 
\prl {\bf 62}, 82 (1989); D.-H. Lee and S.-C. Zhang, \prl {\bf 66}, 1220 (1991). 
%
\bibitem{lee-kane} D.-H. Lee and C.L. Kane, Phys. Rev. Lett. {\bf 64}, 1313
(1990).
%
\bibitem{bose-hubbard} M. P. A. Fisher, P. B. Weichmann, G. Grinstein, and D. S. Fisher, 
\prb {\bf 40}, 546 (1989). 
%
\bibitem{BH-MFT} K. Sheshadri, H. R. Krishnamurthy, R. Pandit, and 
T. V. Ramakrishnan,
\epl {\bf 22}, 257 (1993); E. Altman, W. Hofstetter, E. Demler, 
and M. D. Lukin, \njp {\bf 5}, 113.1 (2003); A. B. Kuklov and
B. V. Svistunov, \prl {\bf 90}, 100401 (2003); A. B. Kuklov,
N. Prokof'ev, and B. V. Svistunov, \prl {\bf 92}, 030403 (2004).
%
\bibitem{Higgs-CS-large-N}  W. Chen, M. P. A. Fisher, and Y.-S. Wu, 
\prb {\bf 48}, 13749 (1993); X.-G. Wen and Y.-S. Wu, \prl {\bf 70}, 
1501 (1993); L. P. Pryadko and S.-C. Zhang, \prb {\bf 54}, 4953
(1996); J. Ye and S. Sachdev, \prl {\bf 80}, 5409 (1996).
%
\bibitem{subir-book} S. Sachdev, {\it Quantum Phase Transitions} 
(Cambridge University Press, New York 1999)
%
\bibitem{spin-glass-QHFM} J. Rapsch, D. K. Lee, and 
J. T. Chalker, \prl {\bf 88}, 036801 (2002); D. K. Lee, S. Rapsch, and
J. T. Chalker, \prb {\bf 67}, 195322 (2003).
%
\bibitem{yang-WC} K. Yang, \prl {\bf 87}, 056802 (2001); J. Ye and L. Jiang, \prl {\bf 98}, 236802 (2007).
%
\bibitem{caldeira} R. L. Doretto, A. O. Caldeira, and 
C. Morais Smith, \prl {\bf 97}, 1876401 (2006);  Z. Papi\'c, and M. V. Milovanovi\'c, \prb {\bf 75}, 195304 (2007). 
%
\bibitem{localization-in-QHFM} S. John and M. J. Stephen, \prb 
{\bf 28}, 6358 (1983); A. G. Green, \prl {\bf 82}, 5104 (1999).
%
\bibitem{disorder-in-QHFM1} A. G. Green, \prb {\bf 57}, R9373 (1998).

\end{thebibliography}
\end{document}